\documentclass[reprint,twocolumn,showpacs,superscriptaddress,longbibliography,email,floatfix,aps,prl]{revtex4-2}

\usepackage{header}

\usepackage{tikz}
\definecolor{CB1}{HTML}{d7191c} 
\definecolor{CB2}{HTML}{fdae61}
\definecolor{CB3}{HTML}{ffffbf}
\definecolor{CB4}{HTML}{abd9e9}
\definecolor{CB5}{HTML}{2c7bb6} 
\definecolor{CB6}{HTML}{32414b} 

\def \xgate{
    \tikz[every picture/.style={line width=1pt}, baseline=-3]{
        
        \draw [every picture] (0,0) -- (1,0);
        
        \filldraw [every picture][fill=CB5!60!white,draw=black] (0.25,-0.25) rectangle (0.75,0.25);
        \node at (0.5,0) {$\varphi_x$};
        \clip (0,-0.25) rectangle (1, 0.25);
        
    }
}

\def \gammagate{
    \tikz[every picture/.style={line width=1pt}, baseline=6]{
    
        \def \spacing{0.6}
    
        \draw [every picture] (0,0) -- (1,0);
        \draw [every picture] (0,\spacing) -- (1,\spacing);
        \draw [every picture] (0.5,\spacing) -- (0.5,0);
        \filldraw [every picture][fill=white,draw=black] (0.5,\spacing) circle (3pt);
        \filldraw [every picture][fill=CB1!60!white,draw=black] (0.25,-0.25) rectangle (0.75,0.25);
        \node at (0.5,0) {$\varphi_\gamma$};
        
        \clip (0,-0.125) rectangle (1, \spacing);
        
    }
}

\def \nngate{
    \tikz[every picture/.style={line width=1pt}, baseline=6]{
    
        \def \spacing{0.6}
    
        \draw [every picture] (0,0) -- (1,0);
        \draw [every picture] (0,\spacing) -- (1,\spacing);
        \filldraw [every picture][fill=CB6!30!white,draw=black] (0.25,-0.125) rectangle (0.75,\spacing + 0.125);
        \node[rotate=-90] at (0.5, 0.5*\spacing) {$\varphi_{nn}$}; 
        
        \clip (0,-0.125) rectangle (1, \spacing);
        
    }
}

\begin{document}
\title{Space-time correlations in monitored kinetically constrained discrete-time quantum dynamics}
\author{Marcel Cech}
\affiliation{Institut f\"ur Theoretische Physik and Center for Integrated Quantum Science and Technology, Universit\"at T\"ubingen, Auf der Morgenstelle 14, 72076 T\"ubingen, Germany}
\author{María Cea}
\affiliation{Max-Plank-Institut f\"ur Quantenoptik, Hans-Kopfermann-Str. 1, D-85748 Garching, Germany}
\affiliation{Munich Center for Quantum Science and Technology (MCQST), Schellingstr. 4, D-80799 M\"unchen, Germany}
\author{Mari Carmen Bañuls}
\affiliation{Max-Plank-Institut f\"ur Quantenoptik, Hans-Kopfermann-Str. 1, D-85748 Garching, Germany}
\affiliation{Munich Center for Quantum Science and Technology (MCQST), Schellingstr. 4, D-80799 M\"unchen, Germany}
\author{Igor Lesanovsky}
\affiliation{Institut f\"ur Theoretische Physik and Center for Integrated Quantum Science and Technology, Universit\"at T\"ubingen, Auf der Morgenstelle 14, 72076 T\"ubingen, Germany}
\affiliation{School of Physics and Astronomy and Centre for the Mathematics and Theoretical Physics of Quantum Non-Equilibrium Systems, The University of Nottingham, Nottingham, NG7 2RD, United Kingdom}
\author{Federico Carollo}
\affiliation{Centre for Fluid and Complex Systems, Coventry University, Coventry, CV1 2TT, United Kingdom}

\begin{abstract}
    State-of-the-art quantum simulators permit local temporal control of interactions and midcircuit readout. These capabilities open the way towards the exploration of intriguing nonequilibrium phenomena. We illustrate this with a kinetically constrained many-body quantum system that has a natural implementation on Rydberg quantum simulators. The evolution proceeds in discrete time and is generated by repeatedly entangling the system with an auxiliary environment that is monitored and reset after each time-step. Despite featuring an uncorrelated infinite-temperature average stationary state, the dynamics displays  coexistence of fast and slow space-time regions in stochastic realizations of the system state. The time-record of measurement outcomes on the environment serves as natural probe for such dynamical heterogeneity, which we characterize using tools from large deviation theory. Our work establishes the large deviation framework for discrete-time open quantum many-body systems as a means to characterize complex dynamics and collective phenomena in quantum processors and simulators.
\end{abstract}

\maketitle


\textbf{Introduction. ---} Statistical mechanics is typically concerned with the analysis of equilibrium and nonequilibrium stationary properties of physical systems \cite{Goldenfeld1992,Ruelle2004,Touchette2009}. Certain many-body systems may however be characterized by ``seemingly trivial", i.e., orderless stationary states while displaying intricate collective dynamical behavior \cite{Lesanovsky2013,Lesanovsky2013a,Lesanovsky2014,Valado2016,Olmos2014,Ferioli2023,DeFazio2024}. Kinetically constrained models provide a paradigmatic example of such interesting (typically slow) dynamics towards structureless thermodynamic equilibrium states \cite{Lecomte2007,Garrahan2009}.
These many-body systems are described by simple dynamical rules that only allow for transitions between single-particle states if a given condition, e.g., the absence or presence of particles in their close neighborhood, is satisfied. This reduction in the connectivity between configurations of the system can result in highly correlated and heterogeneous dynamics  \cite{Cancrini2008,Gutierrez2015,Wilkinson2020,Chen2022,Bertini2024,Gavazzoni2024,Zhang2024}. Within the framework of classical stochastic systems, methods from large deviation theory have been employed to investigate collective dynamical behavior in kinetically constrained models, leading, for instance, to new insights into the glass transition \cite{Ritort2003,Merolle2005,Garrahan2007,Binder2011,Sanders2015,Garrahan2018,Banuls2019,Causer2020,Causer2021,Causer2022}. Extending these models to the quantum realm holds the promise to reveal new dynamical behavior and nonequilibrium phases \cite{Ates2012,Marcuzzi2017,Horssen2015,Lan2018,Gillman2020,Pancotti2020,Magoni2024,Zhang2024}. Interestingly, this undertaking also constitutes a daunting task, in particular due to the intrinsic complexity of many-body quantum systems.

\begin{figure}[t]
    \centering
    \includegraphics{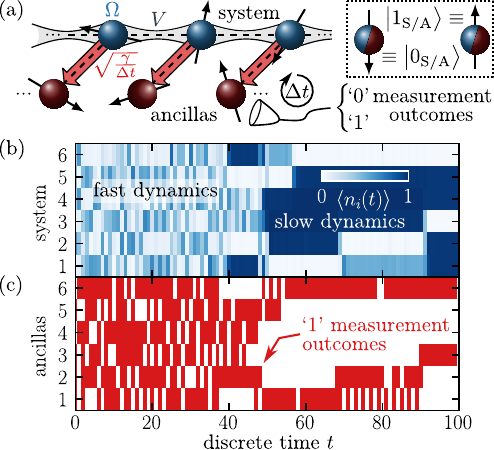}
    \caption{\textbf{Discrete-time dynamics in the Rydberg collision model.} (a)~Collision model setup described by Eqs.~(\ref{eq:collision_hamiltonian},\ref{eq:Rydberg_Hamiltonian}). The interacting quantum system consisting of $L$ two-level systems (blue qubits) is coupled to an ancillary chain (red qubits) of the same length. After the collision time $\Delta t$, the ancillas are measured in their computational basis $\{\ket{0_\mathrm{A}}, \ket{1_\mathrm{A}}\}$ and reset to $\ket{0_\mathrm{A}}$. (b)~Stochastic system dynamics for $L=6$, $\Delta t = 1.25 / \Omega$, $V = 5.875\,\Omega$ and $\gamma = 3\,\Omega$. Iterating the collision model dynamics, the quantum system can show  coexistence of fast and slow dynamical regimes, visible, for instance, in the probability of observing system qubits in $\ket{1_\mathrm{S}}$,  $\expval{n_i(t)}$. (c)~Corresponding quantum trajectory. Remarkably, also the ancillary measurement time-record, here associated with the stochastic realization in panel~(b), shows signatures of the dynamical heterogeneity by displaying distinct space-time regions, e.g., the absence of `1' measurements.}
    \label{fig:fig1}
\end{figure}

Currently available quantum simulation platforms are promising candidates to address this open challenge \cite{Feynman1982,Preskill2018,Preskill2023,ValenciaTortora2022,Koeylueoglu2024,Feldmeier2024,ValenciaTortora2024,Zhao2024b}. Improvements in both the fidelity and the versatility of implementable quantum operations already allowed for a deeper understanding of  quantum many-body dynamics \cite{Richerme2014,Islam2015,Gross2017,Bernien2017,Gross2021,RodriguezVega2022,Huggins2022,Yang2023,Chertkov2023,Guimaraes2023,Kim2023,Mi2024,Fauseweh2024,Cornish2024,Pichard2024}. Most notably, in-circuit measurements and a high level of control over  interactions in neutral atom platforms \cite{Cesa2023,Bornet2024,Bluvstein2024,Ocola2024,Anand2024}, ionic quantum processors \cite{Bruzewicz2019,Pino2021,Heussen2023,Guo2024,Iqbal2024} and superconducting quantum computers \cite{McKay2016,McKay2016,Krantz2019,Alexander2020,Koh2023,Kehrer2024} allow for the direct implementation of engineered discrete-time open-system dynamics \cite{GarciaPerez2020,Cattaneo2021,Cattaneo2022,Strasberg2017,Ciccarello2022,Cilluffo2021,Cech2023,Horssen2015a}. In these settings, open quantum dynamics are realized by sequentially coupling the system with auxiliary degrees of freedom [cf.~Fig.~\ref{fig:fig1}(a)], which are subsequently measured and reinitialized. The time-record of measurements on these {\it ancillary} (or environmental) units provides so-called quantum trajectories, revealing \textit{in-situ} information on the stochastic  dynamics  \cite{Lesanovsky2013a,PerezEspigares2019,Cilluffo2020,Wang2019,Carollo2019a,Horssen2015a,Cilluffo2021,Ciccarello2022,Vovk2022,Cabot2023,Cech2023,Paulino2024a,Brown2024,Eissler2024,Causer2024,Liu2024,Wu2024a,Landi2023,Passarelli2024}. Characterizing the quantum state in single stochastic realizations of the dynamics  [cf.~Fig.~\ref{fig:fig1}(b)] is, however, not possible for many-body systems and for long observation times \cite{Breuer2002,Landi2023,Passarelli2024,Li2024e}. 

In this paper, we propose an approach capable to uncover and systematically characterize distinct dynamical phases in discrete-time many-body quantum dynamics. Our framework is tailored for open systems and solely utilizes information that is efficiently accessible in noisy intermediate scale quantum (NISQ) devices. The analysis is indeed based on so-called ``dynamical order parameters", which can be defined from measurement time-records  [cf.~Fig.~\ref{fig:fig1}(c)] in single dynamical realizations.  
To illustrate our ideas, we consider a model, inspired by the physics of driven-dissipative Rydberg gases \cite{Lesanovsky2013,Lesanovsky2013a,Lesanovsky2014,Valado2016,PerezEspigares2018,Lesanovsky2019}, which features effectively kinetically constrained dynamics in certain parameter regimes \cite{Lesanovsky2012a,Turner2018,Turner2018a,Serbyn2021,Moudgalya2022a}. Despite the average evolution being characterized by a ``seemingly trivial" stationary state, stochastic realizations display complex correlated and heterogeneous dynamics, which also becomes apparent in the corresponding ancillary measurement time-record [see Fig.~\ref{fig:fig1}(b,c)]. 
Exploiting a thermodynamic analogy \cite{Jack2010,Garrahan2010,Chetrite2011,Chetrite2013,Chetrite2015,Carollo2018,Cilluffo2021,Cech2023}, we uncover the presence of macroscopically distinct dynamical regions, which coexist in space and time. This coexistence is responsible for large fluctuations in the monitored dynamics. Our findings widen the use case of Rydberg quantum simulators and quantum computers, by showing their potential for exploring new dynamical phases in monitored open quantum many-body dynamics. \\


\textbf{Collision model setup. ---} We consider discrete-time open quantum dynamics emerging from repeated interactions of the system of interest with ancillary degrees of freedom. The latter encode an effective environment and undergo  measurements and resets at stroboscopic times  \cite{GarciaPerez2020,Horssen2015a,Cattaneo2021,Cattaneo2022,Strasberg2017,Cilluffo2021,Ciccarello2022,Cech2023}. Both the system ($\mathrm{S}$) and the ancillas ($\mathrm{A}$) are described by a chain of $L$ two-level systems, or qubits, with local computational basis $\{\ket{\smash{0_{\mathrm{S}/\mathrm{A}}}}, \ket{\smash{1_{\mathrm{S}/\mathrm{A}}}}\}$ [see Fig.~\ref{fig:fig1}(a)]. A discrete time-step is realized in the following way. The system unitarily collides with the auxiliary chain, initialized in $\ket{\mathbf{0}_\mathrm{A}} = \bigotimes_{i=1}^L \ket{0_\mathrm{A}}_i$. During the interaction, the unitary operator $U = e^{-i H_\mathrm{CM} \Delta t}$, implemented by the Hamiltonian $H_\mathrm{CM}$, entangles system and  ancillas. Afterwards, a projective measurement onto the computational basis is performed on the ancilla qubits. The measurement outcome is collected in a binary string $\mathbf{k} = (k_1, ..., k_L)$, reporting the outcomes $k_i$ for each qubit, and the ancillary chain is then reset to its initial state in preparation for the next collision. The system, for instance initialized in $\ket{\psi(0)} = \ket{\mathbf{0}_\mathrm{S}}$ as well, thus evolves by one discrete-time step according to the relation  $\ket{\psi} \to \ket{\psi'} = K_\mathbf{k} \ket{\psi} / \| K_\mathbf{k} \ket{\psi} \|$, with probability  $\pi(\mathbf{k}) = \| K_\mathbf{k} \ket{\psi} \|^2$. Here, the Kraus operator $ K_\mathbf{k} = \bra{\mathbf{k}_\mathrm{A}}U\ket{\mathbf{0}_\mathrm{A}}$ describes the reduced evolution of the system conditional to the measurement outcome $\mathbf{k} = (k_1, ..., k_L)$ for the ancillary qubits. A single realization of the stochastic dynamics, $\ket{\psi(0)} \to ... \to \ket{\psi(T)}$, is thereby conditioned on the quantum trajectory $\eta(T) = [\mathbf{k}(t)]_{t=1}^T$, i.e., the space-time record of all ancillary measurements [see Fig.~\ref{fig:fig1}(b,c)], and is observed with a probability of $\pi(\eta(T)) = \| K_{\mathbf{k}(T)} ... K_{\mathbf{k}(1)} \ket{\psi(0)}\|^2$. In contrast, the single time-step unconditional average evolution is governed by the Kraus map $\rho\to\rho'=\mathcal{E}[\rho]$, with  $\mathcal{E}[\rho] \coloneqq \sum_{\mathbf{k}} K_\mathbf{k} \rho K_\mathbf{k}^\dagger$ \cite{Kraus1983,Nielsen2010,Cemin2024a}. 

The model we consider is specified by the collision Hamiltonian
\begin{align}
    H_\mathrm{CM} = H_\mathrm{S} \otimes \mathds{1} + \sqrt{\frac{\gamma}{\Delta t}} \sum_{i = 1}^L P_i \otimes {\tau}_i^x \, ,
    \label{eq:collision_hamiltonian}
\end{align}
with $P = \ketbra{0_\mathrm{S}}$ and $\tau^x = \ketbra{0_\mathrm{A}}{1_\mathrm{A}} + \mathrm{h.c.}$, where $\gamma$ represents a dephasing rate in the continuous-time limit $\Delta t \to 0$, and the reset of ancillas can be omitted (see Supplemental Material \cite{SM}\vphantom{\cite{Loew2012,Johansson2013,Wiseman2009,Hatano2005}} for details). The system Hamiltonian $H_\mathrm{S}$ furthermore models a coherent drive with Rabi frequency $\Omega$ and nearest-neighbor interaction with strength $V$, when both qubits are in the (Rydberg) state $\ket{1_\mathrm{S}}$  \cite{Saffman2010,Bernien2017}. We thus define (assuming periodic boundary conditions for the chain)
\begin{align}
    H_\mathrm{S} = \Omega \sum_{i=1}^L {\sigma}_i^x + V \sum_{i=1}^L {n}_i {n}_{i+1} \, ,
    \label{eq:Rydberg_Hamiltonian}
\end{align}
with $\sigma^x = \ketbra{0_\mathrm{S}}{1_\mathrm{S}} + \mathrm{h.c.}$ and $n = \ketbra{1_\mathrm{S}} = 1 - P$. For large interaction strengths $V$, the coherent drive of a qubit located near a qubit in state $\ket{1_\mathrm{S}}$ is highly detuned. In the limit of infinitely strong interactions, $V \to \infty$, this results in the so-called Rydberg-blockade mechanism \cite{Gaetan2009,Wilk2010,Isenhower2010,Zhang2010}, for which sites close to a qubit in state $\ket{1_\mathrm{S}}$ are temporarily frozen. This effect leads to a strict separation of the Hilbert space into sectors labeled by the set of frozen sites with adjacent qubits in state $\ket{1_\mathrm{S}}$  \cite{Lesanovsky2012a,Turner2018,Turner2018a,Serbyn2021,Moudgalya2022a}. In such a limit, the Hamiltonian $H_\mathrm{S}$ can be mapped onto  the so-called PXP Hamiltonian $H_\mathrm{PXP} = \sum_{i=1}^{L} P_{i-1} \sigma_i^x P_{i+1}$, which realizes a kinetically constrained model.  For finite interaction strengths $V$, the Rydberg blockade is not perfect. As such, on short to intermediate time-scales, the Hamiltonian $H_\mathrm{S}$  is well approximated by the PXP Hamiltonian, but the system eventually explores the full Hilbert space at long enough times. 

In Fig.~\ref{fig:fig1}(b), we consider a stochastic realization of the collision model dynamics. Specifically, we show the local probability $\expval{n_i(t)}$ of qubits in the state $\ket{1_\mathrm{S}}$, which immediately allows one to identify two dynamical regimes: a fast regime, which is characterized by rapid changes of $\expval{ n_i(t) }$, and a slow one  where adjacent qubits are close to being in state $\ket{1_\mathrm{S}}$, $\expval{n_i(t)} \approx 1$. Due to the Rydberg-blockade effect, these space-time regions are preserved over several time-steps and substantially slow down the evolution of neighboring qubits as well. It is important to note, that 
the expectation values $\expval{ n_i(t) }$, as shown in Fig.~\ref{fig:fig1}(b), are not directly accessible on a quantum computer. The reason is that for estimating $\expval{ n_i(t) }$ of a stochastic realization, repeated measurements of the operator $n_i(t)$ have to be taken on identical quantum trajectories [cf.~Fig.~\ref{fig:fig1}(c)]. This comes with a postselection overhead that grows exponentially in time and system size making it essentially inaccessible to simulations on a quantum processor \cite{Breuer2002,Landi2023,Passarelli2024,Li2024e}.
However, signatures of the above-mentioned dynamical regimes are already clearly visible in the quantum trajectories themselves, i.e., in the space-time records of the measurement outcomes on the ancillary qubits shown in Fig.~\ref{fig:fig1}(c). The fast regime is then related to ancillas being frequently observed in $\ket{1_\mathrm{A}}$. On the other hand, space-time regions containing adjacent system qubits with $\expval{n_i(t)} \approx 1$, which are found in the slow regime, are reflected in a lack of `1' ancillary measurement outcomes. This connection is a consequence of the structure of the system-ancilla interaction $\propto  P_i \otimes \tau_i^x $ in Eq.~\eqref{eq:collision_hamiltonian},  solely allowing for a rotation of the $i$th ancilla into the state $\ket{1_\mathrm{A}}$ if the corresponding system qubit features a finite overlap with the state $\ket{0_\mathrm{S}}$. The heterogeneous dynamics shown in Figs.~\ref{fig:fig1}(b,c) is even more intriguing when considering that the average stationary state  $\rho_\mathrm{ss}$, i.e., $\mathcal{E}[\rho_\mathrm{ss}] = \rho_\mathrm{ss}$, is the fully mixed state $\mathds{1} / 2^L$ \cite{SM}. \\


\begin{figure}[t]
    \centering
    \includegraphics{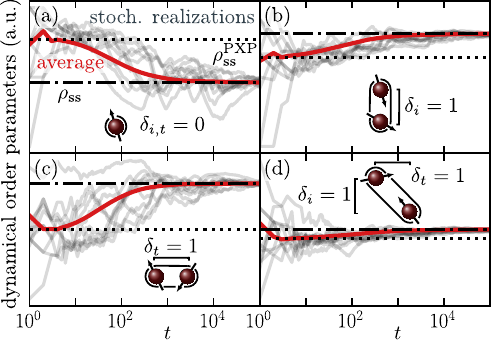}
    \caption{\textbf{Dynamical order parameters for $L = 6$, $\Delta t = 1.25 / \Omega$, $V = 5.875\,\Omega$ and $\gamma = 3\,\Omega$.} (a)~Activity. (b-d)~Space-time correlations with space-time distance ${\bm{\delta}}$. We compare the dynamical order parameters of the ensemble (red lines) with the corresponding observables (see main text) extracted from 10 quantum trajectories (light gray lines). Using Eqs.~(\ref{eq:time_local_prob},\ref{eq:time_nonlocal_prob}), the dash-dotted black line represents the stationary value obtained on $\rho_\mathrm{ss} = \mathds{1} / 2^L$, while the dotted black line contains the prediction of the projection onto the free sector of the PXP model $\rho_\mathrm{ss}^\mathrm{PXP} \propto \mathcal{P}_\mathrm{PXP}[\rho_\mathrm{ss}]$ with no adjacent qubits in state $\ket{1_\mathrm{S}}$. }
    \label{fig:fig2}
\end{figure}

\textbf{Space-time correlations in quantum trajectories. ---}  We characterize dynamical behavior in  quantum trajectories by means of so-called dynamical order parameters. For the purpose of this work these are the {\it activity}
\begin{align}
    a(t) = \frac{1}{L} \sum_{i=1}^{L} \overline{k_i(t)} \, 
    \label{eq:activity}
\end{align}
and the {\it space-time correlations} defined through the spatially averaged covariance
\begin{align}
    c_{\bm{\delta}}(t) \! = \! \frac{1}{L} \sum_{\mathclap{i = 1}}^{L} \!\Big[ \overline{k_{i}(t \!-\! \delta_t) k_{i+\delta_i}(t)} - \overline{k_{i}(t \!-\! \delta_t)} \, \overline{k_{i+\delta_i}(t)} \Big]\, .
    \label{eq:correlations}
\end{align}
The former observable keeps track of the frequency of $k_i(t) = 1$ measurement outcomes, while the latter one of their correlations when outcomes are considered at the space-time distance $\bm{\delta} = (\delta_i, \delta_t)$. The average $\overline{\,\cdot\,}$ in Eqs.~(\ref{eq:activity},\ref{eq:correlations}) is an ensemble average over all possible quantum trajectories, weighted with their respective probabilities. In terms of the average state $\rho(t) = \mathcal{E}^t[\rho(0)]$, the relevant probabilities for computing the  ensemble averages in Eqs.~(\ref{eq:activity},\ref{eq:correlations}) are
\begin{align}
    p(\mathbf{k}(t)\!=\!\mathbf{k}) &= \Tr{{K}_{\mathbf{k}} \rho(t\!-\!1) {K}_{\mathbf{k}}^\dagger} \, ,
    \label{eq:time_local_prob}
\end{align}
for single-time and 
\begin{align}
    p(\mathbf{k}(t)\!=\!\mathbf{k}, \mathbf{k}(t\!-\!1)\!=\!\mathbf{k}') &= \Tr{{K}_{\mathbf{k}} {K}_{\mathbf{k}'} \rho(t\!-\!2) {K}_{\mathbf{k}'}^\dagger {K}_{\mathbf{k}}^\dagger} \, ,
    \label{eq:time_nonlocal_prob}
\end{align}
for two-time functions \cite{Landi2023}. The above equations show that, asymptotically in time,  dynamical order parameters are completely described by the stationary state $\rho_\mathrm{ss}$.

The calculation of the above quantities still requires sampling multiple dynamical realizations on quantum processors. Interestingly, we can define closely related observables, displaying analogous behavior, which are solely defined on individual quantum trajectories $\eta(T) = [\mathbf{k}(t)]_{t=1}^T$. These time-integrated observables are of the following form 
\begin{align}
    \mathcal{O}_{\bm{\delta}}(\eta(T)) = \sum_{\mathclap{t = 1+\delta_t}}^{\mathclap{T}} ~ \sum_{\mathclap{i=1}}^L k_i(t - \delta_t) k_{i + \delta_i}(t) \, .
    \label{eq:trajectory_observables}
\end{align}
The rationale is that time averages and ensemble averages coincide at stationarity. This allows us to compare $\mathcal{O}_{\mathbf{0}}(\eta(T)) / {(LT)}$ and $\mathcal{O}_{\bm{\delta}}(\eta(T)) /  (LT') - [\mathcal{O}_{\mathbf{0}}(\eta(T)) /  (LT)]^2$, where $T' =  (T-\delta_t)$ denotes the extensivity of the first sum in Eq.~\eqref{eq:trajectory_observables},
with the activity and the space-time correlations in Eqs.~(\ref{eq:activity},\ref{eq:correlations}), respectively. 

In Fig.~\ref{fig:fig2}, we test this understanding by comparing the activity and space-time correlations with the corresponding observables in Eq.~\eqref{eq:trajectory_observables} calculated for 10 quantum trajectories. We furthermore show the prediction of the stationary state $\rho_\mathrm{ss} = \mathds{1} / 2^L$ and of its projection onto the free PXP sector, $\rho_\mathrm{ss}^\mathrm{PXP} \propto \mathcal{P}_\mathrm{PXP}[\rho_\mathrm{ss}]$, with no adjacent qubits in state $\ket{1_\mathrm{S}}$, as selected by the initial state $\ket{\psi(0)} = \ket{\mathbf{0}_\mathrm{S}}$ \cite{Turner2018,Li2023}. We observe that, for a long (metastable) timescale, dynamical order parameters are initially described by the prediction obtained with $\rho_\mathrm{ss}^\mathrm{PXP}$. After this transient regime, the system overcomes the blockade effect, which is not perfect due to finite values of $V$, and the average state eventually approaches $\rho_\mathrm{ss} = \mathds{1} / 2^L$. The  sampled quantum trajectories and their time-integrated observables follow the same trend and converge towards the same stationary values. We observe two important  differences. First, we note a delay in the transient behavior of the quantum trajectories over the instantaneous ensemble-averaged dynamical order parameters. 
This delay is expected since time-integrated observables have some ``inertia'', due to the fact that measurement probabilities depend on the current state  $\pi(\mathbf{k}(t)\!=\!\mathbf{k})\!=\!\| K_\mathbf{k} \ket{\psi(t\!-\!1)}\|^2$ and due to the time average over the whole measurement history.
Additionally, these quantities show a large variance over long time-periods. Recalling the dynamical heterogeneity observed in Figs.~\ref{fig:fig1}(b,c),  these fluctuations are due to the intermittent exploration of different PXP sectors, based on the position of adjacent qubits in state $\ket{1_\mathrm{S}}$. With the system being approximately confined in one sector over an extended period of time, observations dominantly account for the single-sector contribution. For very long times, the system eventually explores the Hilbert space uniformly (see $\rho_\mathrm{ss} = \mathds{1} / 2^L$) and also dynamical order parameters of single quantum trajectories approach the stationary prediction. \\


\begin{figure}[t]
    \centering
    \includegraphics{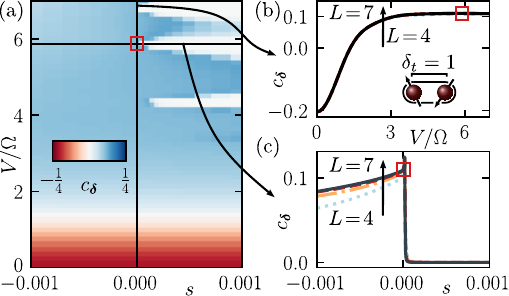}
    \caption{\textbf{Dynamical phase diagram.} (a)~Temporal correlations $c_{\bm{\delta}}$, with $\bm{\delta} = (0, 1)$, for $L = 6$, $\Delta t = 1.25 / \Omega$, $\gamma = 3\,\Omega$. The phase diagram in the interaction strength $V$ and counting field $s$ displays dynamical phases with decreased activity ($s > 0$) that potentially come with vanishing temporal correlations. For $s = 0$ and $L = 4$ to $L = 7$, panel~(b) shows that these distinct dynamical phases do not affect the bare average of dynamical observables. However, the sharp change with $s$ displayed in panel~(c) for $V = 5.875\,\Omega$, reveals nearby dynamical phases with drastically different temporal correlations that result in the large fluctuations observed Fig.~\ref{fig:fig2}.
    Note that we show the dynamical phase diagram of the respective ensembles at stationarity (see Supplemental Material~\cite{SM} for details).}
    \label{fig:fig3}
\end{figure}

\textbf{Unraveling dynamical phases. ---} 
The observed dynamical heterogeneity and large fluctuations are reminiscent of first-order phase transitions and phase coexistence  \cite{Goldenfeld1992}. In what follows, we make this connection   concrete by means of a large deviation approach \cite{Touchette2009}. We consider the ensemble of quantum trajectories generated by the collision model, $\{\eta(T),\pi(\eta(T))\}$, and interpret this as a microcanonical ensemble. We further introduce a counting field $s$, playing the role of an inverse temperature, associated with a time-integrated observable $\mathcal{O}$, embodying instead an energy function \cite{Garrahan2010,Cilluffo2021,Cech2023,Horssen2015a,Carollo2018}. Through these quantities, we define a canonical ensemble with probabilities $\pi(\eta(T),s)\propto e^{-s\mathcal{O}(\eta(T))}\pi(\eta(T))$. By varying the counting field $s$, we can thus control the probability over quantum trajectories (see Supplemental Material~\cite{SM} for details on the large deviation approach as well as its potential implementation on a quantum device).
As we will show, this thermodynamic-like construction allows us to compute a suitable phase diagram for the stochastic process, demonstrating that the observed phenomenology is related to the coexistence of dynamical phases.

Based on the observation in Fig.~\ref{fig:fig1} that slow dynamics are accompanied by a space-time region of primarily `0' ancillary measurement outcomes, we take the time-integrated activity $\mathcal{O}_\mathbf{0}$ as observable associated with the counting field. Therefore, positive (negative) values of $s$ will decrease (increase) the  typical activity in the canonical ensemble. 
We consider the temporal correlations $c_{\bm{\delta}}$, with $\bm{\delta} = (0, 1)$, as our dynamical order parameter and investigate its stationary behavior upon variations of the counting field $s$ and of the interaction strength $V$. The resulting phase diagram is shown in Fig.~\ref{fig:fig3}(a). 
At first, we observe temporal correlations that slowly increase with the interaction strength. Furthermore, for $s > 0$, i.e., for decreasing activities, we observe a drastic decrease of temporal correlations for certain interaction strengths $V$. Taking a closer look at $s = 0$, see Fig.~\ref{fig:fig3}(b), we see that the dynamical order parameter changes smoothly with $V$. This fact can be understood by recalling Eqs.~(\ref{eq:time_local_prob},\ref{eq:time_nonlocal_prob}) and considering that the stationary state, for $s=0$, is the fully mixed state for any interaction strength $V$. 
Conversely, Fig.~\ref{fig:fig3}(c) shows that temporal correlations can suddenly drop for certain values of $V$, upon considering low-activity regimes ($s>0$).   This abrupt change in the behavior of the dynamical order parameter in the vicinity of $s \approx 0$, which is the regime associated with collision model dynamics, explains the observed heterogeneity [see Fig.~\ref{fig:fig1}(b,c)]  as stemming from the coexistence of different dynamical phases. This thus connects the observed phenomenology to the emergence of a first-order transition in the dynamical phase diagram of the collision model. It further demonstrates how large deviation methods allow to unravel complex dynamical behavior in discrete-time open quantum dynamics.

\textbf{Conclusions and outlook. --- } We have investigated discrete-time open-system dynamics inspired by the dissipative evolution of driven Rydberg atoms. Despite having a seemingly trivial stationary state, the model we have considered displays dynamical heterogeneity, stemming from the coexistence of fast and slow dynamics in single stochastic realizations. We have shown how such collective behavior can be characterized by means of dynamical order parameters, such as the frequency and the correlations of measurement outcomes on the environmental ancillary qubits. The investigation performed here is limited to relatively small system sizes, which however already show clear signatures of many-body phenomena (see Supplemental Material \cite{SM} for the direct analysis of fluctuations as a function of the system size). In order to address significantly larger system sizes, it would be necessary to devise suitable tensor-network approaches \cite{Yang2020,Gillman2021,Fux2023,Causer2024}. 
Such larger systems should be, however, already accessible on state-of-the-art quantum processors, offering an intriguing perspective for the exploration of dynamical heterogeneity in glassy relaxation.

The code and the data that support the findings of this work are available on Zenodo \cite{ZenodoData}.\\


\acknowledgments
\textbf{Acknowledgements. --- } We thank Dominik Sulz, Cecilia De Fazio and Farokh Mivehvar for fruitful discussions. We acknowledge funding from the Deutsche Forschungsgemeinschaft (DFG, German Research Foundation) under Germany's Excellence Strategy -- EXC-2111 -- 390814868, through the Research Unit FOR 5413/1, Grant No. 465199066, and through the Research Unit FOR 5522/1, Grant No. 499180199. This project has also received funding from the European Union’s Horizon Europe research and innovation program under Grant Agreement No. 101046968 (BRISQ). F.C.~is indebted to the Baden-W\"urttemberg Stiftung for the financial support of this research project by the Eliteprogramme for Postdocs.

\bibliography{biblio}

\onecolumngrid
\clearpage


\setcounter{equation}{0}
\setcounter{page}{1}

\setcounter{figure}{0}
\setcounter{table}{0}
\makeatletter
\renewcommand{\theequation}{S\arabic{equation}}
\renewcommand{\thefigure}{S\arabic{figure}}
\renewcommand{\thetable}{S\arabic{table}}
\setcounter{secnumdepth}{1}

\begin{center}
{\Large SUPPLEMENTAL MATERIAL}
\end{center}
\begin{center}
\vspace{0.8cm}
{\Large Space-time correlations in monitored kinetically constrained discrete-time quantum dynamics}
\end{center}
\begin{center}
Marcel Cech,$^{1}$ Mar\'ia Cea,$^{2, 3}$ Mari Carmen Ba\~nuls,$^{2, 3}$ Igor Lesanovsky,$^{1,4}$ and Federico Carollo$^5$ 
\end{center}
\begin{center}
$^1${\em Institut f\"ur Theoretische Physik and Center for Integrated Quantum Science and Technology, Universit\"at T\"ubingen, Auf der Morgenstelle 14, 72076 T\"ubingen, Germany}\\
$^2${\em Max-Plank-Institut f\"ur Quantenoptik, Hans-Kopfermann-Str. 1, D-85748 Garching, Germany}\\
$^3${\em Munich Center for Quantum Science and Technology (MCQST), Schellingstr. 4, D-80799 M\"unchen, Germany}\\
$^4${\em School of Physics and Astronomy and Centre for the Mathematics and Theoretical Physics of Quantum Non-Equilibrium Systems, The University of Nottingham, Nottingham, NG7 2RD, United Kingdom}\\
$^5${\em Centre for Fluid and Complex Systems, Coventry University, Coventry, CV1 2TT, United Kingdom}
\end{center}


\section{Continuous-time dynamics}
In this section, we connect the continuous-time limit, $\Delta t \to 0$, of the collision model dynamics with a Lindblad master equation describing driven and interacting Rydberg atoms subject to local dephasing. We then study the phase diagram, when we control the activity in the canonical ensemble. 

\begin{figure}[ht]
    \centering
    \includegraphics{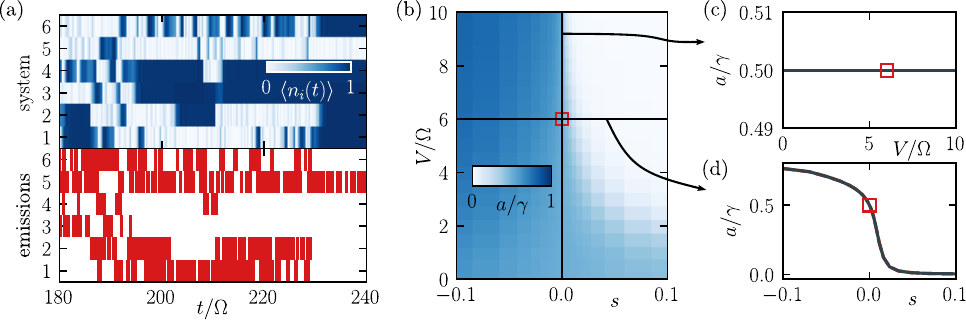}
    \caption{\textbf{Continuous-time dynamics. } (a) Stochastic realization and quantum trajectories for $L = 6$, $V = 6 \,\Omega$ and $\gamma = 3 \, \Omega$. In the top panel, we show the spatially resolved probability $\expval{n_i(t)}$ to find a qubit in state $\ket{1_\mathrm{S}}$. In the bottom panel, a red stroke at the space-time $(i,t)$ represents an emission event associated with the action of the jump operator $J_i = \sqrt{\gamma} P_i$ at time $t$. (b) Dynamical phases for $L=6$ qubits and $\gamma = 3 \, \Omega$. We change the interaction strength $V$ and the conjugate field $s$ and plot the stationary activity in the respective ensemble. (c) Cut at $s = 0$. As outlined in the text, we observe a constant stationary activity $a_\mathrm{ss} = \gamma / 2$, when changing $V$. (d) Cut at $V = 6 \, \Omega$. We show the dependence of the stationary activity on the counting field $s$ for $L = 4$ to $L = 7$. The curves for different system sizes overlap, but the separation of dynamical phases is not as sharp as in Fig.~\ref{fig:fig3}.}
    \label{fig:fig_s1}
\end{figure}

In a first step, we perform the continuous-time limit $\Delta t \to 0$. The average collision model dynamics, with $U = e^{-i \Delta t (H_\mathrm{S} \otimes \mathds{1} + H_\mathrm{int})}$ [see Eq.~\eqref{eq:collision_hamiltonian}], of system in state $\rho(t)$ for small collision times and ancillas initialized in $\rho_\mathrm{A} =\ketbra{\mathbf{0}_\mathrm{A}}$ can be written as 
\begin{align}
    \frac{\rho(t) - \rho(t\!-\!1)}{\Delta t} \approx -i \Big[ H_\mathrm{S} + \Tr_\mathrm{A}\{H_\mathrm{int} \rho_\mathrm{A}\}, \rho(t\!-\!1) \Big] + \sum_{\{\mathbf{k}\}} \left(J_\mathbf{k} \rho(t\!-\!1) J_\mathbf{k}^\dagger - \frac{1}{2} \left\{J_\mathbf{k}^\dagger J_\mathbf{k}, \rho(t\!-\!1) \right\} \right)\, .
    \label{eq:small_collision_times}
\end{align}
In Eq.~\eqref{eq:small_collision_times}, $\Tr_\mathrm{A}$ denotes the trace over all ancillary two-level systems, $\{X, Y\} = XY + YX$ the anticommutator and $J_\mathbf{k} = \bra{\mathbf{k}_\mathrm{A}} H_\mathrm{int} \ket{\mathbf{0}_\mathrm{A}} {\sqrt{\Delta t}}$ the potential jump-operators \cite{Ciccarello2022}. Based on Eq.~\eqref{eq:collision_hamiltonian}, the system-ancilla interaction is $H_\mathrm{int} = \sqrt{\gamma / \Delta t} \left( \sum_{i=1}^L P_i \otimes \tau_i^x \right)$ and thus, we obtain $\Tr_\mathrm{A}\{H_\mathrm{int} \rho_\mathrm{A}\} = 0$ and $J_\mathbf{k} = \delta_{\mathbf{k}, \mathbf{e}_i} \sqrt{\gamma} P_i$, where $\delta_{\mathbf{k}, \mathbf{e}_i}$ denotes the componentwise Kronecker-delta of $\mathbf{k}$ and the $i$th standard basis vector $\mathbf{e}_i$. Therefore, the Lindblad master equation
\begin{align}
    \dot{\rho} = \mathcal{L}[\rho] = -i \left[ H_\mathrm{S}, \rho \right] + \sum_{i=1}^L \left( J_i \rho J_i - \frac{1}{2} \left\{ J_i^2, \rho \right\} \right) \, ,
    \label{eq:continuous_time_Lindblad}
\end{align}
with Hamiltonian $H_\mathrm{S} = \Omega \sum_{i=1}^L {\sigma}_i^x + V \sum_{i=1}^L {n}_i {n}_{i+1}$ and hermitian jump operators $J_i = \sqrt{\gamma} P_i$, describes the continuous-time limit. Here, $H_\mathrm{S}$ models driven Rydberg atoms, whose interactions are approximated by the dominant nearest-neighbor contribution, and $\{J_i\}$ resembles the intrinsic or engineered dephasing \cite{Loew2012,Valado2016,PerezEspigares2018,Lesanovsky2013}. 

In Fig.~\ref{fig:fig_s1}(a), we now investigate the dynamics by means of quantum jump trajectories obtained by unraveling the master equation with the quantum Hamiltonian $H_\mathrm{S}$ and jump operators $\{ J_i\}$. The simulations are performed using the python package QuTiP and its Monte-Carlo solver \texttt{mcsolve} \cite{Johansson2013}. In the top panel, we show the spatially resolved probability $\expval{n_i(t)}$ to find the $i$th qubit in $\ket{1_\mathrm{S}}$. Similar to Fig.~\ref{fig:fig1}, we observe that adjacent qubits in this state, if dynamically created, show slower changes of this probability than individual ones. This heterogeneity is also reflected in the lower panel, where we plot the locally resolved emissions, i.e., actions of the respective jump operators $\{ J_i\}$. Here, the space-time region of adjacent qubits with $\expval{n_i(t)} \approx 1$ in the system is completely inactive. 

Similar to the approach in the main manuscript, we consider the instantaneous activity $a(t)$ associated with frequency of abrupt changes per qubit in the quantum jump Monte-Carlo approach. On average, the activity is given by $a(t) = ( \sum_{i = 1}^L \Tr{J_i^2 \rho(t)} ) / L$ \cite{Breuer2002,Wiseman2009}. Since Eq.~\eqref{eq:continuous_time_Lindblad} features  hermitian jump operators $J_i$, the stationary state is again the fully mixed state $\rho_\mathrm{ss} = \mathds{1} / 2^L$. Moreover, we can directly evaluate the stationary activity and find $a_\mathrm{ss} = \gamma / 2$, since $\Tr{J_i^2}=\gamma 2^{L-1}$. 

In order to access the fluctuations in the activity, we resort again to the thermodynamics of quantum trajectories \cite{Garrahan2010,Carollo2018}. Here, we calculate the dominant eigenvalue of the tilted Lindblad master equation
\begin{align}
    \mathcal{L}_s[\rho] = -i[H_\mathrm{S}, \rho] + \sum_{i = 1}^L \left(e^{-s} J_i \rho J_i - \frac{1}{2} \left\{ J_i^2, \rho \right\}\right) \, ,
\end{align}
to extract the scaled cumulant generating function $\theta(s)$ of the activity. Analogously to the approach in the main manuscript, the counting field $s$ controls the activity of the canonical ensemble that we construct utilizing a Gibbs weight with the time-integrated activity embodying an energy function. Panel~(b) shows the first scaled cumulant, $-\theta'(s)$, which describes the phase diagram of the stationary activity. Utilizing the cuts at $s=0$ in panel~(c) and at $V = 6\,\Omega$ in panel~(d), we observe that the heterogeneity in panel~(a) can be understood in terms of two distinct dynamical phases. While for $s \leq 0$, the activity is (approximately) constant in $V$, it decreases for positive counting fields $s$ as we increase $V$. However, we also observe that the separation between the distinct dynamical phases, i.e., inactive at $s > 0$ and active at $s < 0$, is not as sharp as in the discrete-time dynamics considered in the main manuscript, for similar values of $V/\Omega$. 


\section{Unitality and stationary state}
As anticipated in the main manuscript, the quantum channel representing the average dynamics considered in this work is unital, i.e., $\mathcal{E}[\mathds{1}] = \mathds{1}$. In this section, we provide the corresponding arguments for this statement and moreover argue that the fully mixed state $\mathds{1} / 2^L$ is the stationary state of the average dynamics for finite interaction strengths $V$.

We start with the observation that the collision Hamiltonian in Eq.~\eqref{eq:collision_hamiltonian} only contains hermitian and commuting operators $\tau_i^x$ that act on the ancillas. Furthermore, we utilize the freedom of choice in the trace to write 
\begin{align}
    \mathcal{E}[\rho] = \Tr_\mathrm{A}\{ U (\rho \otimes \ketbra{\mathbf{0_\mathrm{A}}}) U^\dagger \} = \sum_{\{m\}} \bra{m_\mathrm{A}} U \ket{\mathbf{0}_\mathrm{A}} \rho \bra{\mathbf{0}_\mathrm{A}} U^\dagger \ket{m_\mathrm{A}} = \sum_{\{m\}} K_m \rho K_m^\dagger \, ,
\end{align}
with potentially different Kraus operators $K_m = \bra{m_\mathrm{A}} U \ket{\mathbf{0}_\mathrm{A}}$ for the chosen basis enumerated by $m$. Note that this simply establishes the unitary freedom of Kraus operators representing a certain quantum channel \cite{Nielsen2010}. Choosing the eigenbasis labeled by $m$ such that each $\ket{m_\mathrm{A}}$ is an eigenstate of the ancillary part of the system-ancilla interaction, we immediately see that every Kraus operators $K_m$ becomes proportional to a unitary matrix. Therefore, together with the trace-preservation of the quantum channel $\mathcal{E}$, i.e., $\mathcal{E}^*[\mathds{1}] = \sum_{\{m\}} K_m^\dagger K_m = \mathds{1}$, we find that $\mathcal{E}[\mathds{1}] = \sum_{\{m\}} K_m K_m^\dagger = \mathds{1}$.

Finally, we argue that the fully mixed state is the stationary state $\rho_\mathrm{ss}$ of the average dynamics for finite $V$. Unlike for the PXP model, the finite interaction strength $V$ and the transverse driving with $\Omega$ ensure that there is no sector of the Hilbert space that is dynamically disconnected from any other. The unitality therefore makes the fully mixed state $\mathds{1} / 2^L$ the (unique) stationary state of the dynamics. 


\section{\label{sec:reset_free_implementation}Reset-free implementation}
In this section, we outline a simple postprocessing scheme that enables the implementation of the discrete-time open-system dynamics described by Eq.~\eqref{eq:collision_hamiltonian} without resetting the ancillas after every collision.

In contrast to the arguments to prove that the quantum channel is unital, the unraveling into stochastic realizations depends genuinely on the specific choice of the measurement basis. For the specific collision model in this work however, we notice that the system-ancilla interaction in Eq.~\eqref{eq:collision_hamiltonian} is inversion symmetric on the ancillary part, i.e., invariant under the exchange of $\ket{0_\mathrm{A}}$ and $\ket{1_\mathrm{A}}$. We can exploit this symmetry to investigate closer $K_{\mathbf{k}|\mathbf{k}'} = \bra{\mathbf{k}_\mathrm{A}} U \ket{\mathbf{k}'_\mathrm{A}}$, representing the observation of the ancilla measurement $\mathbf{k}$ after previously observing $\mathbf{k}'$. By omitting the reset, $K_{|\mathbf{k} - \mathbf{k}'|} = \bra{| \mathbf{k} - \mathbf{k}'|_\mathrm{A}} U \ket{\mathbf{0}_\mathrm{A}}$ in the usual collision model setup is replaced by the Kraus operator $K_{\mathbf{k}|\mathbf{k}'}$ in the reset-free implementation. 
Therefore, reinterpreting a change in the measurement outcome of the $i$th ancilla between two time-steps in the reset-free quantum trajectory stands in one-to-one correspondence with the observation of a `1' measurement outcome in the usual collision model setup that resets the ancillas before the collision takes place.


\section{Probabilities and dynamical order parameters}
In this section, we discuss the necessary steps to  obtain Eqs.~(\ref{eq:time_local_prob},\ref{eq:time_nonlocal_prob}) and demonstrate their application in deriving  the instantaneous activity and space-time correlations in Eqs.~(\ref{eq:activity},\ref{eq:correlations}). For most of this discussion, we follow the steps outlined in Ref.~\cite{Landi2023} that directly apply to our collision model dynamics. 

Directly from the definition of the stochastic realizations in discrete-time open quantum dynamics, the probability to observe a specific trajectory is given by
\begin{align}
    \pi(\eta(T) \!=\! [\mathbf{k}(1), ..., \mathbf{k}(T)]) = \left\| K_{\mathbf{k}(T)} ... K_{\mathbf{k}(1)} \ket{\psi(0)}\right\|^2 
    = \Tr{K_{\mathbf{k}(T)} ... K_{\mathbf{k}(1)} \rho(0) K_{\mathbf{k}(1)}^\dagger ... K_{\mathbf{k}(T)}^\dagger} \, ,
\end{align}
with the initial density matrix $\rho(0) = \ketbra{\psi(0)}$. To calculate the instantaneous activity and space-time correlations in Eqs.~(\ref{eq:activity},\ref{eq:correlations}), we marginalize over measurement outcomes that do not appear in the considered discrete times. For example, the probability to observe the measurement outcomes $\mathbf{k}$ at the discrete-time $t$ is  
\begin{align}
    \begin{split}
        p(\mathbf{k}(t)\!=\!\mathbf{k}) &= \sum_{\{\eta\}} \pi(\eta\!=\![\mathbf{k}(1), ..., \mathbf{k}(t)]) \, \delta_{\mathbf{k}(t),\mathbf{k}}\\
        &= \sum_{\{\mathbf{k}(t-1)\}} ... \sum_{\{\mathbf{k}(1)\}} \Tr{K_{\mathbf{k}} {K_{\mathbf{k}(t-1)} ... K_{\mathbf{k}(1)} \rho(0) K_{\mathbf{k}(1)}^\dagger ... K_{\mathbf{k}(t-1)}^\dagger} K_{\mathbf{k}}^\dagger} \\
        &= \Tr{K_{\mathbf{k}} \rho(t\!-\!1) K_{\mathbf{k}}^\dagger} \, ,
    \end{split}
\end{align}
where we utilize the subsequent application of Kraus maps as in $\rho(1) = \sum_{\{\mathbf{k}(1)\}} K_{\mathbf{k}(1)} \rho(0) K_{\mathbf{k}(1)}^\dagger$. Analogously, we find the joint probability of $\mathbf{k}$ and $\mathbf{k}'$ as $\mathbf{k}(t)$ and $\mathbf{k}(t\!-\!\delta_t)$ to be
\begin{align}
    p(\mathbf{k}(t)\!=\!\mathbf{k}, \mathbf{k}(t\!-\!\delta_t)\!=\!\mathbf{k}') &= \Tr{{K}_{\mathbf{k}} \mathcal{E}^{\delta_t-1}\left[{K}_{\mathbf{k}'} \rho(t\!-\!\delta_t\!-\!1) {K}_{\mathbf{k}'}^\dagger\right] {K}_{\mathbf{k}}^\dagger} \, .
    \label{eq:time_nonlocal_prob_derivation}
\end{align}
In this way, we obtain both Eq.~\eqref{eq:time_local_prob} and a generalized expression of Eq.~\eqref{eq:time_nonlocal_prob} for $\delta_t \geq 1$. Note that we introduce a prime in the above notation to better distinguish the measurement results at $t\!-\!\delta_t$ and at $t$ in Eqs.~(\ref{eq:time_nonlocal_prob},\ref{eq:time_nonlocal_prob_derivation}). 

We can utilize these probabilities to obtain the instantaneous activity
\begin{align}
    \overline{k_i(t)} = \sum_{\{\mathbf{k}\}} p(\mathbf{k}(t)\!=\!\mathbf{k}) \, k_i \, ,
    \label{eq:time_local_connection}
\end{align}
and space-time correlations
\begin{align}
    \overline{k_{i}(t \!-\! \delta_t) k_{i+\delta_i}(t)} = \sum_{\{\mathbf{k}, \mathbf{k}'\}} p(\mathbf{k}(t)\!=\!\mathbf{k}, \mathbf{k}(t\!-\!\delta_t)\!=\!\mathbf{k}') \, k_{i\!+\!\delta_i} k'_i \, ,
    \label{eq:time_nonlocal_connection}
\end{align}
as the weighted sums of the ensemble's probabilities. 


\section{Unraveling dynamical phases with biased many-body dynamics \label{seq:biased_many_body_dynamics}}
In this section, we outline the calculations to obtain the dynamical phase diagram in Fig.~\ref{fig:fig3} of the main manuscript. We therefore connect the canonical ensemble constructed from the counting field $s$ and the time-integrated activity $\mathcal{O}_\mathbf{0}$ with an auxiliary collision model dynamics \cite{Cech2023}. In particular, we focus on the long-time limit, where the dynamics essentially coincides with the time-independent biased dynamics in Ref.~\cite{Cilluffo2021}.

We briefly recall the canonical ensemble of quantum trajectories. Its probabilities are defined as
\begin{align}
    \pi(\eta(T), s) = \frac{1}{Z_T(s)} e^{-s \mathcal{O}_\mathbf{0}(\eta(T))} \pi(\eta(T)) \, ,
    \label{eq:biased_probabilities}
\end{align}
where 
\begin{align}
    Z_T(s) = \sum_{\{\eta(T)\}} e^{-s \mathcal{O}_\mathbf{0}(\eta(T))} \pi(\eta(T))
    \label{eq:partition_function}
\end{align}
is a normalization that we neglected for brevity in the main manuscript. Since $e^{-s \mathcal{O}_\mathbf{0}(\eta(T))}$ implements a bias of the trajectory's probability that depends on the number of observed `1' measurement outcomes, we notice that auxiliary dynamics are essentially implemented by the tilted Kraus operators $e^{-s  \mathcal{O}_\mathbf{0}(\mathbf{k}) / 2} K_\mathbf{k}$. In this expression, $\mathcal{O}_\mathbf{0}(\mathbf{k})$ counts the amount of `1' measurement outcomes in $\mathbf{k}$. Note however that the hereby defined tilted Kraus map $\mathcal{E}_s[\rho] = \sum_{\{\mathbf{k}\}} e^{-s  \mathcal{O}_\mathbf{0}(\mathbf{k})} K_\mathbf{k} \rho K_\mathbf{k}^\dagger$ is no physical quantum channel as it violates trace preservation for $s \neq 0$. Therefore, it does not permit consistent probabilities in Eqs.~(\ref{eq:time_local_prob},\ref{eq:time_nonlocal_prob}) to evaluate Eqs.~(\ref{eq:activity},\ref{eq:correlations}), but generates the normalization $Z_T(s) = \Tr{\mathcal{E}_s^T[\rho(0)]}$.

As demonstrated in Ref.~\cite{Cech2023}, we can define a generalized rotation to obtain physical dynamics with the intended properties by means of time-dependent biased Kraus operators 
\begin{align}
    \tilde{K}_{\mathbf{k}(t)} = e^{-s \mathcal{O}_\mathbf{0}(\mathbf{k}) / 2} G_t K_\mathbf{k} G_{t-1}^{-1} \, ,
\end{align}
with the recursion relation $G_{t-1} = \sqrt{\mathcal{E}_s^*[(G_t)^2]}$. Here, the dual tilted Kraus map $\mathcal{E}_s^*$ acts sequentially on the operator $G_t$ backpropagating it to earlier times. Asymptotically, these operators are described in terms of the dominant matrix $l_s$ of the dual tilted Kraus map $\mathcal{E}_s^*$. The corresponding eigenvalue equation yields $\mathcal{E}_s^*[l_s] = \Lambda_s l_s$. Therefore, the time-independent biased Kraus operators 
\begin{align}
    \tilde{K}_\mathbf{k} = \frac{e^{-s \mathcal{O}_\mathbf{0}(\mathbf{k}) / 2}}{\Lambda_s^{1/2}} l_s^{1/2} K_\mathbf{k} l_s^{-1/2} 
    \label{eq:biased_Kraus_operators}
\end{align}
emerge \cite{Cilluffo2021}. The biased Kraus operators in Eq.~\eqref{eq:biased_Kraus_operators} implement \textit{bona-fide} quantum dynamics. In particular, the completely positive trace-preserving quantum channel $\tilde{\mathcal{E}}[\rho] \coloneqq \sum_{\mathbf{k}} \tilde{K}_\mathbf{k} \rho \tilde{K}_\mathbf{k}^\dagger$ allows for a proper probabilistic interpretation, while encoding the probabilities in the canonical ensemble at large times.
In particular, we obtain the dynamical phase diagram (see, e.g., Fig.~\ref{fig:fig3}) by first calculating the stationary state $\tilde{\rho}_\mathrm{ss}$ of the biased dynamics, which is usually different from $\rho_\mathrm{ss}$ for $s \neq 0$, before applying Eqs.~(\ref{eq:time_local_prob},\ref{eq:time_nonlocal_prob}). Using Eqs.~(\ref{eq:time_local_connection},\ref{eq:time_nonlocal_connection}), the corresponding dynamical phase diagram represents the stationary behavior, i.e., $T \to \infty$, upon variations of $s$. \\

\begin{figure}
    \centering
    \includegraphics{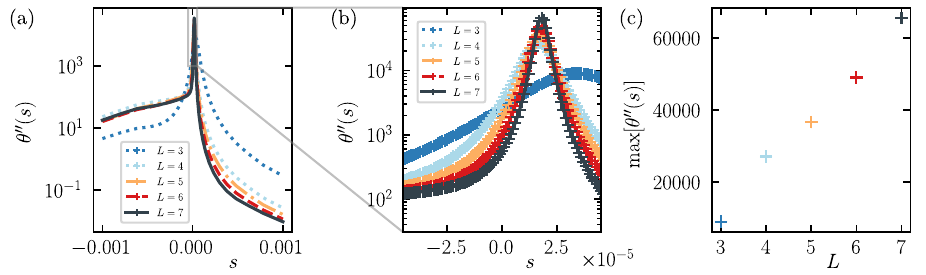}
    \caption{\textbf{Unravelling dynamical phases with regards to dynamical fluctuations.} (a)~Second-order scaled cumulant of the time-integrated activity $\mathcal{O}_\mathbf{0}$ obtained during the calculation of biased Kraus operators in Fig.~\ref{fig:fig3}(c). (b)~Zoom into the $s$ parameter region with the largest $\theta''(s)$ value. (c)~Maximum of the second-order scaled cumulant as a function of the size of the system $L$. For all evaluations, we consider system sizes $L=3$ to $L=7$ with $\Delta t = 1.25 / \Omega$, $V = 5.875\,\Omega$ and $\gamma = 3\,\Omega$.}
    \label{fig:fig_s2}
\end{figure}

Within this approach (for example, when computing the dynamical phase diagram in Fig.~\ref{fig:fig3}), we also obtain the scaled cumulant generating function 
\begin{align}
    \theta(s) = \lim_{T\to\infty} \frac{1}{LT} \log Z_T(s)
    = \frac{1}{L} \log(\Lambda_s)
\end{align}
for the time-integrated activity $\mathcal{O}_\mathbf{0}$ \cite{Cilluffo2021}. By taking the derivative with respect to the counting field $s$, we obtain the scaled cumulants
\begin{align}
    \kappa^{(1)}(s) &= -\theta'(s) = \frac{1}{LT} {\mathcal{O}_\mathbf{0}}(T, s) \quad \equiv \text{mean}(s)\, , \\
    \kappa^{(2)}(s) &= \theta''(s) = \frac{1}{LT} \left({\mathcal{O}_\mathbf{0}^2}(T, s) - ({\mathcal{O}_\mathbf{0}}(T, s))^2 \right)\quad \equiv \text{scaled variance}(s)\, ,
    \label{eq:second_derivative_theta_s}
\end{align}
in the canonical ensemble specified by $s$, where the biased probabilities $\pi_s(\eta(T))$ in Eq.~\eqref{eq:biased_probabilities} enter $\mathcal{O}(T, s) = \sum_{\{\eta(T)\}} \pi_s(\eta(T)) \mathcal{O}(\eta(s))$. 
Resembling the discontinuity of the order parameter in a first-order phase transition \cite{Touchette2009}, the scaled cumulant generating function $\theta(s)$ in the thermodynamic limit $L \to \infty$ is non-differentiable at the critical counting field $s_c$. 
The discontinuous order parameter should hence be signalled by a diverging second-order scaled cumulant at $s_c$ as the system size $L$ is increased. 

With this consideration in mind, we plot the scaled variance in Fig.~\ref{fig:fig_s2}(a) as a function of $s$. Here, we notice that its value increases significantly in the vicinity of $s\approx 0$. 
Investigating closer this region in Fig.~\ref{fig:fig_s2}(b), we observe that its peak, which is actually slightly shifted from $s=0$, becomes more and more pronounced as $L$ increases. Considering the maximum of the scaled variance for different system sizes reveals a linear increase with $L$ as shown in Fig.~\ref{fig:fig_s2}(c). 
While we cannot access larger system sizes, the large absolute value of the scaled variance signals the dynamical heterogeneity observed in Fig.~\ref{fig:fig1} and the  large fluctuations in Fig.~\ref{fig:fig2}.


\section{Quantum simulation of the collision model dynamics}
In this section, we discuss how the collision-model dynamics investigated in this work can be implemented on quantum simulators and computers. To this end, we also simulate postselection-free ancilla-measurement time-records which may be obtained directly from an quantum processor and explore the resulting dynamical phase diagram as in Fig.~\ref{fig:fig3}. 

\begin{figure}
    \centering
    \includegraphics{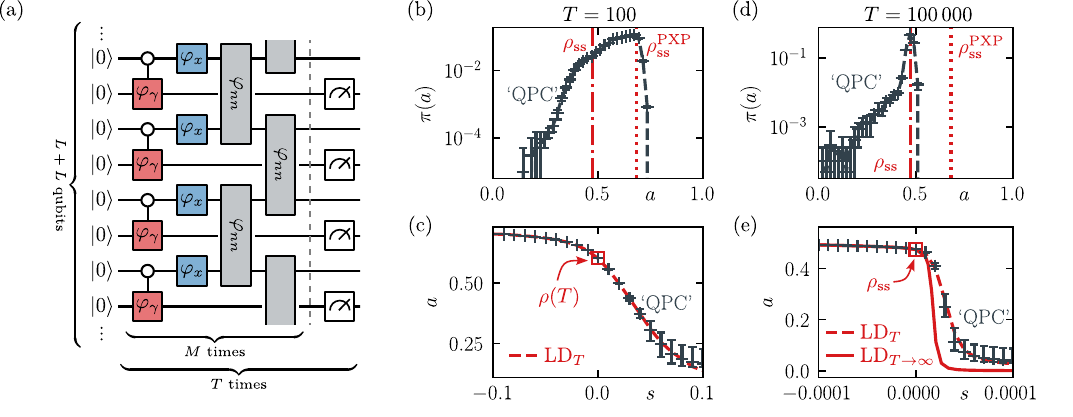}
    \caption{\textbf{Quantum simulation of the collision model dynamics.} (a)~Quantum circuit representation. Using the Trotter decomposition in Eq.~\eqref{eq:trotter_decomposition} and the corresponding gates in Eq.~\eqref{eq:Trotter_gates}, we define a quantum circuit to implement the Rydberg collision model on a digital quantum computer.
    The quantum trajectory of one experimental run is directly encoded in the $L \times T$ mid-circuit measurements.
    (b)~Histogram of the activity $a = \mathcal{O}_\mathbf{0}(\eta(T)) / (LT)$ after $T = 100$ collisions for $L=6$ with $\Delta t = 1.25 / \Omega$, $V = 5.875\,\Omega$ and $\gamma = 3\,\Omega$. We compare the simulation results from $100\,000$ quantum trajectories under ideal conditions [gray crosses] with the predictions obtained on the stationary state $\rho_\mathrm{ss}$ [red dash-dotted line] and on its projection onto the free sector of the PXP model $\rho_\mathrm{ss}^\mathrm{PXP} \propto \mathcal{P}_\mathrm{PXP}[\rho_\mathrm{ss}]$ with no adjacent qubits in state $\ket{1_\mathrm{S}}$ [red dotted line]. 
    (c)~Activity phase diagram. Using the simulated experimental runs in panel~(b), we calculate the activity with respect to the biased empirical probabilities in Eq.~\eqref{eq:biased_probabilities} [gray crosses]. For comparison, we also show the prediction of the average state $\rho(T)$ [red square] as well as the finite time predictions that stem from the average state evolution under $\mathcal{E}_s$ [red dashed line]. 
    (d,e)~Histogram and dynamical phase diagram of the activity for $1000$ simulated quantum trajectories with $T = 100\,000$ each, analogous to panels~(b) and (c). 
    For reference, we also show the prediction for the stationary, i.e., $T \to \infty$, limit of the activity phase diagram by utilizing the biased many-body dynamics in Eq.~\eqref{eq:biased_Kraus_operators} [red solid line].
    The error bars in panels~(b-e) are obtained by resampling all simulated quantum trajectories into five separate data sets and calculating their spread.
    }
    \label{fig:fig_s3}
\end{figure}

Let us first note that the unitary dynamics with mid-circuit measurements of our bipartite system-ancilla dynamics is within reach of current capabilities of  dual-species Rydberg quantum simulator, as recently discussed in Ref.~\cite{Anand2024}. Engineering the inter- and intra-species interactions alongside a global laser drive of the system atoms might therefore directly implement the collision model as described through Eqs.~(\ref{eq:collision_hamiltonian},\ref{eq:Rydberg_Hamiltonian}).

Alternatively, we can approximate the joint evolution via a quantum circuit that is then executed by a digital  quantum computer [see Fig.~\ref{fig:fig_s3}(a)].
We can approximate the dynamics exploiting a first-order Trotter decomposition to obtain \cite{Hatano2005} 
\begin{align}
    U = e^{-i H_\mathrm{CM} \Delta t} = \left(e^{-i H_\mathrm{CM} \Delta t / M}\right)^M 
    \approx 
    \left(  
        \underbrace{
        e^{-i\frac{V \Delta t}{M} \sum n_i n_{i+1}} \cdot
        e^{-i\frac{\Omega \Delta t}{M} \sum \sigma_i^x}
        }_\text{system only} \cdot
        \underbrace{
        e^{-i\frac{\sqrt{\gamma \Delta t}}{M} \sum P_i \otimes \tau_i^x }
        }_\text{system-ancilla int.}
    \right)^M\, .
    \label{eq:trotter_decomposition}
\end{align}
Utilizing the reset-free implementation with the simple postprocessing of recorded ancilla measurements discussed in section \ref{sec:reset_free_implementation}, we can realize the collision-model dynamics through the circuit depicted in Fig.~\ref{fig:fig_s3}(a). Here, the $L+L$ qubits are initialized according to $\ket{\psi(0)} = \ket{\mathbf{0_\mathrm{S}}}$ and the usual ancilla reference state $\ket{\mathbf{0_\mathrm{A}}}$. 
The gates are defined as 
\begin{align}
    \gammagate = e^{-i \varphi_\gamma (P \otimes \tau^x)}\, , \quad 
    \xgate = e^{-i \varphi_x \sigma^x} \quad \text{and} \quad 
    \nngate = e^{-i \varphi_{nn} (n \otimes n)}\,, \quad 
    \label{eq:Trotter_gates}
\end{align}
with the corresponding angles set to $\varphi_\gamma = \sqrt{\gamma \Delta t} / M$, $\varphi_x = \Omega \Delta t / M$ and $\varphi_{nn} = V \Delta t / M$. A discrete-time step is then completed by the projective measurements in the computational basis $\{\ket{0_\mathrm{A}}, \ket{1_\mathrm{A}}\}$ of the ancilla qubits.

For both implementations, each experimental run returns a particular  time-record of measurements of the ancillas, that is, the quantum trajectory $\eta(T)$ of a stochastic realization. Notably, unlike the properties of the quantum state of the system conditioned on this measurement time-record, time-integrated observables of these measurements [see Eq.~\eqref{eq:trajectory_observables}] do not suffer from a postselection barrier \cite{Breuer2002,Landi2023,Passarelli2024,Li2024e}.

To showcase the potential of realizing our dynamics on a quantum simulator, we  consider an ideal quantum processor implementing the joint time-evolution of system and ancillas via $U = e^{-i H_\mathrm{CM} \Delta t}$. 
We simulate $100\,000$ experimental runs and extract the time-integrated number of `1' ancilla measurement-outcomes  $\mathcal{O}_\mathbf{0}(\eta(T))$ for $T=100$ and $T=100\,000$ collisions. 
In both cases, we investigate the histogram of the activity $a = \mathcal{O}_\mathbf{0}(\eta(T)) / (LT)$ and its corresponding activity phase diagram defined through the biased probabilities in Eq.~\eqref{eq:biased_probabilities}.

Starting with $T=100$ collisions, we observe that the histogram of the activity in Fig.~\ref{fig:fig_s3}(b) shows a broad dome which includes both the prediction of the stationary state $\rho_\mathrm{ss} = \mathds{1} / 2^L$ and its projection onto the free PXP sector, $\rho_\mathrm{ss}^\mathrm{PXP} \propto \mathcal{P}_\mathrm{PXP}[\rho_\mathrm{ss}]$. While the latter is related to the free PXP sector chosen by the initial state $\ket{\psi(0)} = \ket{\mathbf{0}_\mathrm{S}}$, the broad distribution shows that the different dynamical regimes can already be detected in a small number of collisions. In Fig.~\ref{fig:fig_s3}(c), this observation is also captured in the activity phase diagram, which shows a noticeable crossover when reducing the activity for $s>0$.
Moreover, we find that the finite time predictions utilizing $- \frac{1}{L} \frac{\partial}{\partial s} \log{Z_T(s)}$, with $Z_T(s) = \Tr{\mathcal{E}_s^T[\rho(0)]}$ (see Sec.~\ref{seq:biased_many_body_dynamics}), agree well with the biased empirical trajectories.

Moving on to $T=100\,000$ collisions, the histogram in Fig.~\ref{fig:fig_s3}(d) shows a dominant peak around the stationary value of the activity predicted by $\rho_\mathrm{ss} = \mathds{1} / 2^L$. 
The initial state is no longer relevant as we approach stationarity.
The overall shape of the distribution has also changed, and trajectories with significantly fewer `1' ancilla measurements are also sampled. Considering the logarithmic y-scale, the deviation from a Gaussian distribution highlights the apparent, atypical fluctuations.
In fact, these small activity fluctuations are effectively enhanced in Eq.~\eqref{eq:biased_probabilities} for $s>0$, and therefore give rise to the sharp change near $s\approx 0$ of the activity phase diagram shown in Fig.~\ref{fig:fig_s3}(e).
We also make two interesting observations. First, we again find a very good agreement between the 
empirically biased quantum trajectories and the finite time prediction using the average state evolution under $\mathcal{E}_s$. Second, we notice a quantitative difference between the cases with a finite number of collisions and the limit of $T \to \infty$ obtained from the stationary state $\tilde{\rho}_\mathrm{ss}$ of the biased dynamics. It is perfectly clear that finite times can only probe an approximation of the limit with infinitely many collisions needed to fully generate the low-activity ensemble of quantum trajectories with the reweighting in Eq.~\eqref{eq:biased_probabilities}. At the same time, we can appreciate the advantages of the biased many-body dynamics in Eq.~\eqref{eq:biased_Kraus_operators} if they can be computed \cite{Carollo2018}. 


\section{Single-body dynamics}

In this section, we investigate the case of a noninteracting, i.e., single-body, dynamics, achieved by setting $V = 0$ in Eq.~\eqref{eq:Rydberg_Hamiltonian}. We calculate analytically the activity and temporal correlations at stationarity and before exploring the influence of a static detuning $\Delta$ in the system Hamiltonian $H_\mathrm{S}$.

\begin{figure}[ht]
    \centering
    \includegraphics{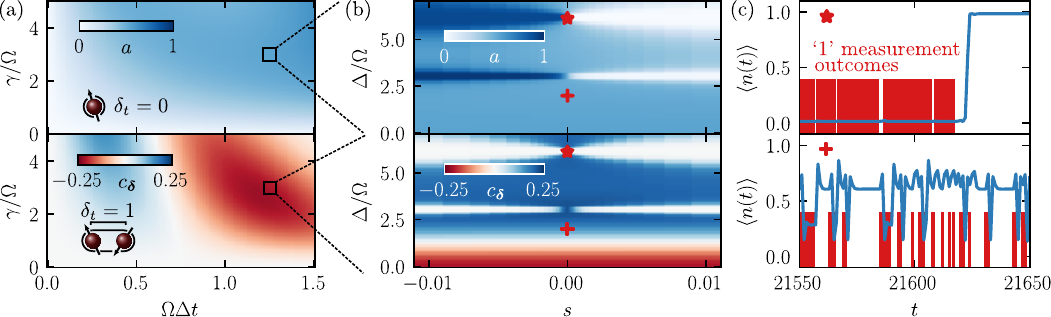}
    \caption{\textbf{Single-body dynamics.} (a)~Dynamical order parameters across $(\Delta t, \gamma)$-plane. For $L=1$, we show the stationary activity $a$ (top) and temporal correlations $c_{\bm \delta}$, with ${\bm \delta} = (0, 1)$ (bottom). (b)~Influence of static detuning. For the parameters $(\Delta t, \gamma) = (1.25 / \Omega, 3 \, \Omega)$, we plot the dependence of both dynamical order parameters on the static detuning $\Delta$ and the activity counting field $s$. (c)~Stochastic realizations. For two chosen detunings $\Delta$ [see star and cross in panel~(b)], we investigate the probability $\expval{n(t)}$ to find the qubit in state $\ket{1_\mathrm{S}}$ (blue line) and the corresponding ancilla measurements (red strokes) of the stochastic realization. }
    \label{fig:fig_s4}
\end{figure}

We start by parametrizing the collision model Hamiltonian in $U = e^{-i H_\mathrm{CM} \Delta t}$ for a single qubit as
\begin{align}
    H_\mathrm{CM} \Delta t = a (\sigma^x \otimes \mathds{1}) + b (P \otimes \tau^x) \, ,
\end{align}
where we define $a = \Omega \Delta t$ and $b = \sqrt{\gamma \Delta t}$. Utilizing the stationary state of the dynamics $\rho_\mathrm{ss} = \mathds{1} / 2$, the stationary probability to observe $k(t) = 1$ reads 
\begin{align}
    p(k(t)\!=\!1) \overset{t\to\infty}{\to} \Tr{K_1 \rho_\mathrm{ss} K_1^\dagger} = \frac{1}{2} \Tr{K_1 K_1^\dagger} = \frac{1}{2} - \frac{1}{2c} \cos b \left(b^2 \cos \sqrt{c}+4 a^2\right) \, ,
    \label{eq:single_body_activity}
\end{align}
with $c = 4 a^2+b^2$. Similarly, in order to also resolve temporal correlations, we compute the probability for two consecutive `1' measurements in time, which is given by
\begin{align}
    \begin{split}
        p&(k(t)\!=\!1,k(t\!-\!1)\!=\!1) \overset{t\to\infty}{\to} \Tr{(K_1)^2 \rho_\mathrm{ss} (K_1^\dagger)^2} = \frac{1}{2} \Tr{(K_1)^2 (K_1^\dagger)^2}\\
        &= \frac{4 a^2 \left(4 b^2 \cos \sqrt{c} \!+\!c \cos (2 b)\right)-4 c \cos b \left(b^2 \cos \sqrt{c}\!+\!4 a^2\right)+b^2 \left(c \cos (2 b)\!-\!4 a^2\right) \cos \left(2 \sqrt{c}\right)+3 \left(16 a^4\!+\!4 a^2 b^2\!+\!b^4\right)}{8 c^2} \, .
    \end{split}
\end{align}
Since here $\overline{k(t)} = p(k(t)\!=\!1)$ and $\overline{k(t) k(t-1)} = p(k(t)\!=\!1,k(t\!-\!1)\!=\!1)$, the activity at stationarity is thus given by Eq.~\ref{eq:single_body_activity}, while the corresponding temporal correlations, with ${\bm{\delta} = (0, 1)}$, read
\begin{align}
    c_{\bm{\delta}}(t) = \overline{k(t) k(t-1)} - \overline{k(t)}\, \overline{k(t-1)} \overset{t \to \infty}{\to}  \frac{1}{2
   c^2} \left( b^2 \sin ^2b \sin ^2\left(\frac{1}{2} \sqrt{c}\right) \left(\left(8 a^2+b^2\right) \cos \sqrt{c}+b^2\right) \right) \, .
\end{align}

We show the explicit dependence of both dynamical order parameters on the collision time $\Delta t$ and dephasing rate $\gamma$ in Fig.~\ref{fig:fig_s4}(a). We observe both temporal correlations and anti-correlations. In particular, for the collision parameters $(\Delta t, \gamma) = (1.25 / \Omega, 3\,\Omega)$ that we utilize throughout our investigations, we find temporal anti-correlations [see also $V = 0$ in Fig.~\ref{fig:fig3}].

For these parameters, we now consider a static detuning $\Delta$ of the $\ket{1_\mathrm{S}}$ state that models an off-resonant Rabi drive. The corresponding system Hamiltonian then yields
\begin{align}
    H_\mathrm{S} = \Omega \, \sigma^x + \Delta \, n \, .
\end{align}

In Fig.~\ref{fig:fig_s4}(b), we plot the activity and temporal correlation phase diagrams for canonical ensembles with $s$ being the counting field with respect to the time-integrated number of `1' measurement outcomes. We observe that both quantities vary smoothly with the detuning $\Delta$ at $s = 0$. This is not surprising, since the fully mixed state $\rho_\mathrm{ss} = \mathds{1} / 2$ remains the stationary state of the dynamics. Conversely, both dynamical order parameters show sharp changes, when varying $s$ at certain recurrent detunings $\Delta$. Here, negative counting fields promote a very active phase, reducing temporal correlation to almost zero, while positive $s$ values lead to very inactive dynamics with also negligible temporal correlations. In Fig.~\ref{fig:fig_s4}(c), we investigate the influence of the sharp changes in dynamical order parameters on stochastic realizations. We show exemplary dynamics for detunings with and without distinct dynamical phases in their immediate vicinity in $s$ [corresponding to the star and cross in the phase diagrams]. In the first case, displayed in the upper panel, we observe intermittent periods of the active and the inactive phase alongside the system qubit being in the $\ket{0_\mathrm{S}}$ state or the $\ket{1_\mathrm{S}}$ state. Conversely, in the lower panel at a detuning, where no distinct dynamical phases meet at $s \approx 0$, the state of the system is not pinned to one of the computational basis states. Furthermore, the measurements are not bunched over a long period of time. 


\end{document}